\documentclass[aps,twocolumn,showpacs]{revtex4}

\usepackage{graphicx}
\usepackage{epsfig}
\usepackage{epstopdf}
\usepackage{amsfonts}
\usepackage{amsmath,amssymb}

\begin{document}

\title{Generalized Richardson-Gaudin Nuclear Models}
\author{J.~Dukelsky$^{1}$,
V.G.~Gueorguiev\footnote{
On leave of absence from the Institute of Nuclear Research and Nuclear
Energy, Bulgarian Academy of Sciences.}$^{1,2}$
and P. Van Isacker$^{3}$}
\address{
$^{1}$Instituto de Estructura de la Materia,
CSIC. Serrano 123, 28006 Madrid, Spain \\
$^{2}$Department of Physics and Astronomy,
Louisiana State University, Baton Rouge, USA\\
$^{3}$Grand Acc\'el\'erateur National d'Ions Lourds,
BP 55027, F-14076 Caen Cedex 5, France}

\date{\today}

\begin{abstract}
The exact solvability of several nuclear models
with non-degenerate single-particle energies is outlined
and leads to a generalization
of integrable Richardson-Gaudin models,
like the $su(2)$-based fermion pairing,
to any simple Lie algebra.
As an example,
the $so(5)\sim sp(4)$ model of $T=1$ pairing is discussed
and illustrated for the case of $^{64}$Ge
with non-degenerate single-particle energies.
\end{abstract}

\maketitle

Exactly solvable models (ESM) built upon a dynamical symmetry have
a long history of providing important insights into the structure
of nuclei. The two main advantages of ESM are: (1) They can
describe in an analytical or exact numerical way a wide variety of elementary phenomena.
(2) They can be and have been used as a testing ground
for various many-body approaches. The simplest example of ESM is
the rank-one $su(2)$ model of fermions in one orbit or in several
degenerate orbits with a constant pairing interaction, often used
to introduce nuclear superconductivity (see e.g.,
Ref.~\cite{superconductivity}).

In general, a quantum system has a dynamical
symmetry if the Hamiltonian can be expressed in terms of the
Casimir operators of a chain of nested algebras. A typical example
of a model with rank-two Lie algebra dynamical symmetry is Elliott's $su(3)$
model, introduced to describe the phenomenon of nuclear
deformation~\cite{Ell58}. Elliott's Hamiltonian is a linear
combination of the quadratic Casimir operator of the $su(3)$ Lie
algebra (involving a quadrupole-quadrupole interaction) and of the
Casimir of its $so(3)$ subalgebra, associated with angular
momentum. Lie algebras with higher rank lead to more complex ESM
like the $so(5)$ model of $T=1$, isovector pairing~\cite{Flo52},
the $so(8)$ model of $T=0,1$ isoscalar and isovector
pairing~\cite{Flo64,Eva81}, Ginocchio's model with $so(8)$ and
$sp(6)$ structure~\cite{Gin80}, also known as the Fermion
Dynamical Symmetry Model (FDSM)~\cite{Wu87}. The three dynamical
symmetries of the Interacting Boson Model (IBM)~\cite{Iac87}
provide another example of this approach.

The concept of quantum integrability goes beyond the limits of the
dynamical symmetry approach. A quantum system is integrable if there exist
as many commuting Hermitian operators (integrals of motion) as
quantum degrees of freedom \cite{Zhang&Feng}. The set of Casimir
operators of a chain of nested algebras fulfills this condition.
There are, however, also well-known examples of integrable models
without dynamical symmetry \cite{none-DinSym-int-mod}.

Usually dynamical symmetry models are defined for degenerate
single-particle levels. Changes in the
single-particle energies break the dynamical symmetry but may
still preserve integrability. The pairing model with
non-degenerate single-particle levels, with an exact solution
found by Richardson in the sixties, represents a unique example of
an ESM with such characteristics~\cite{Ric63}. Recently,
Richardson's model has been shown to be integrable by finding the
complete set of integrals of motion constructed in terms of the
generators of the $su(2)$ algebra~\cite{Pairing-integrability}.
Subsequently, more general exactly solvable pairing models, both
for fermions and for bosons, called the Richardson-Gaudin (RG)
models, have been proposed~\cite{Duk01}. Since then, a great deal
of work has been devoted to understand the properties of these ESM
and to apply them to a wide variety of problems in nuclear,
condensed-matter, atomic, and molecular physics~\cite{DUKE}.

The aim of this Letter is to present the generalization of the RG
models to those symmetry algebras
that have given rise to the well-known ESM in
nuclear physics. In most cases this generalization will allow
non-degenerate single-particle energies, as well as other symmetry
breaking one-body operators. As an example, we shall discuss the
exact solution of the $so(5)$ proton-neutron (pn) isovector pairing
model with non-degenerate orbits, for which an exact solution has
been proposed by Richardson~\cite{Ric}. However, it has been
recently shown that the Richardson's solution was
incorrect~\cite{Pan02}.

We also mention that $so(5)$ has been proposed as the symmetry
underlying high $T_{\rm c}$-superconductivity~\cite{Z1}. The
exactly solvable RG models discussed in this Letter could be used
to generalize $so(5)$ condensed-matter models~\cite{S1} by the
explicit {\it addition} of non-degenerate single-particle
symmetry-breaking terms.

We begin by introducing a set of commuting operators,
the RG operators (integrals of motion) $R_i$~\cite{Aso02}:
\begin{equation}
R_i=\sum_{j(\neq i)}\frac{X_i\cdot X_j}{z_j-z_i}+\xi_i,
\label{Rop}
\end{equation}
where the index $i$ ($j$) refers to the $i$-th ($j$-th) copy
of a Lie algebra $\mathcal{L}$ with generators $X^\alpha_i$, $z_{i}$ are fixed parameters which later will be related to the single-particle energies,
and $\xi$ is a generic element of the Cartan subalgebra of $\mathcal{L}$
(see below).
The scalar product is defined
through the $\mathcal{L}$-invariant metric tensor
$g_{\alpha\beta}=c_{\alpha\rho}^\sigma c_{\beta\sigma}^\rho$
where $c_{\alpha\beta}^\gamma$
are the structure constants of $\mathcal{L}$,
e.g., $X_i\cdot X_j\equiv X^\alpha_i g_{\alpha\beta} X^\beta_j$.
Since the $R_i$ commute,
any function of these operators can be used
as a model Hamiltonian for an integrable system.
In particular, any linear combination of the $R_i$  is integrable and is
at most quadratic in the generators.

For any simple algebra $\mathcal{L}$,
the eigenvalues of the $R_i$ have been given in Ref.~\cite{Aso02}
or can be derived from the Gaudin-algebra approach~\cite{Ush94}:
\begin{equation}
r_i=\Lambda_i\cdot\xi+
\sum_{j(\neq i)}\frac{\Lambda_i\cdot\Lambda_j}{z_j-z_i}+
\sum_{a=1}^r\sum_{\alpha=1}^{M^a}
\frac{\Lambda_i\cdot\pi^a}{z_i-e_{a,\alpha}},
\label{regv}
\end{equation}
where the parameters $e_{a,\alpha}$ are solutions of the
generalized Richardson equations~\cite{Aso02,Ush94}
\begin{equation}
\sum_{b=1}^r \left.\sum_{\beta=1}^{M^b}\right. ^\prime
\frac{\pi^b\cdot\pi^a}{e_{b,\beta}-e_{a,\alpha}}- \sum_{i=1}^L
\frac{\Lambda_i\cdot\pi^a}{z_i-e_{a,\alpha}}=
\xi\cdot\pi^a\equiv\rho^a.
\label{gReq}
\end{equation}
The primed sum means that the singular term $(a,\alpha)=(b,\beta)$ is omitted
and $L$ is the number of copies of the algebra $\mathcal{L}$ of rank $r$.
The Cartan subalgebra $\mathcal{L}^{\rm C}\subset\mathcal{L}$
has the elements $h1,\dots,h^r$
and the $\rho^{a}$ are the components of $\xi$
in this generally non-orthonormal basis.
The $\pi_s^a$ are the components
of the $r$ simple roots $\pi1,\dots,\pi^r$ of $\mathcal{L}$
in the Cartan-Weyl basis, $[h_s,E^a]=\pi_s^aE^a$.
Each $\Lambda_i$ is a vector of highest weight
(usually of the fundamental representation)
for the $i$-th copy of $\mathcal{L}$
and the expression $\Lambda_i\cdot\pi^a$
corresponds to the eigenvalue of $H^a=\sum_s\pi_s^ah^s$
in this highest-weight state of the $i$-th copy of $\mathcal{L}$.
The $\rho^a$ are, in a sense, the strength components
of the symmetry-breaking one-body operator $\xi$.
Finally, the $M^a$ are positive numbers
related to the eigenvalues $m^a$ of $H^a$
in the eigenstate of the integrals of motion~(\ref{Rop}), $M^a=\sum_i(\Lambda_i\cdot\pi^a-m_i^a)$.

Equations~(\ref{regv}) and~(\ref{gReq})
were derived by Asorey {\it et al.}~\cite{Aso02}
under the assumption that $\mathcal{L}$ is the Lie algebra
of a simple, simply-connected, compact Lie group.
We arrived at the same equations
within Ushveridize's framework~\cite{Ush94}
which assumes that $\mathcal{L}$
is a {\it singular} semi-simple Lie algebra
(i.e., a classical algebra
or one of the exceptional algebras $E_6$ or $E_7$).
Since Eqs.~(\ref{regv}) and~(\ref{gReq}) involve a dot product,
well defined for any Lie algebra,
one may expect that the equations themselves
are valid for any semi-simple Lie algebra
and possibly for an even larger class.
In fact, it is desirable to find a formulation of these equations
that involves arbitrarily chosen Cartan algebra generators
instead of the simple roots of the algebra.

Although one can use any function of the $R_i$ as a Hamiltonian,
the following particular linear combination yields a simple
expression for the eigenvalues:
\begin{equation}
H=\sum_iz_iR_i=
\sum_iz_i\xi_i-
\frac{1}{2}\left(C_2-\sum_iC_2^{(i)}\right).
\label{Ham1}
\end{equation}
With the same linear combination of the eigenvalues $r_i$~(\ref{regv})
and using the Richardson equations~(\ref{gReq}),
one can show that the eigenvalues of~(\ref{Ham1})
are linear in the spectral parameters $e_{a,\alpha}$
with coefficients $\delta^a$
that depend on the symmetry-breaking strengths $\rho^a$.

The Hamiltonian~(\ref{Ham1}) shows how the global
$\mathcal{L}$-symmetry represented
by the total second-order Casimir operator $C_2$
is broken by the first term containing the elements $\xi_i$
of the Cartan subalgebras $\mathcal{L}^{\rm C}_i$.
In the general case of $z_i\neq z_j$ the
first term does not commute with $C_2$. Thus, the eigenstates
of~(\ref{Ham1}) are spread over different irreducible
representations of $\mathcal{L}$ and they are not eigenstates
of the total Casimir operator $C_2$.

Using Cartan's classification of semi-simple Lie algebras,
one can now generalize many nuclear physics models
within this framework (see Table~\ref{Table1}).
\begin{table}[htbp]
\caption{Algebras associated with some nuclear physics models;
$\sim$ denotes isomorphisms.
R\&R is the symplectic model of
Rosensteel and Rowe~\cite{Ros77}.
$^{\star}$FDSM Lie algebras \cite{Wu87}.}
\label{Table1}
\begin{ruledtabular}
\begin{tabular}{||c|c|c|c|c||}
rank $n$ & $su(n+1)$ & $so(2n+1)$ & $sp(2n)$ & $so(2n)$\\ \hline
1 & pairing & $\sim su(2)$ & $\sim su(2)$ & $\sim u(1)$\\ \hline
2 & Elliott & $T=1$ pairing& $\sim so(5)$ & $\sim su(2)\oplus su(2)$\\ \hline
3 & Wigner & $so(7)^{\star}$ & R\&R$^{\star}$
& $\sim su(4)^{\star}$\\ \hline
4 & $su(5)$ & $so(9)$ & $sp(8)$ & $T=0,1$ pairing$^{\star}$\\ \end{tabular}
\end{ruledtabular}
\end{table}
Note that models based on a fermion realization of the type
$X_{\alpha\beta}\sim a_\alpha^{\dagger}a_\beta^{\dagger}$, such as
$sp(2n)$ or $so(n)$, are well adapted for the generalization
towards non-degenerate single-particle energies because in that
case the Cartan generators are sums of fermion number operators.
In contrast, models based on realizations of the form
$a_\alpha^{\dagger}a_\beta$, such as $su(n)$, contain differences
of number operators.

To illustrate the present formalism, we shall discuss the example
of $T=1$ pairing in systems with protons and neutrons and
non-degenerate single-particle levels. The dynamical symmetry
limit of pn-pairing in a single degenerate shell was first
considered by Flowers~\cite{Flo52} as an important extension of
the original pairing problem between identical particles. It
yields a seniority classification of protons and neutrons in $jj$
coupling in terms of pp, nn, and pn pairs. In doing so, the
concepts of seniority $v$ (the number of nucleons not in pairs
coupled to angular momentum zero) and of reduced isospin $t$ (the
isospin of these nucleons) are established. The application of
this $so(5)$ formalism has given rise to many analytic reduction
formulae for shell-model matrix elements~\cite{Hec67}.

For simplicity, in the RG extension of the $so(5)$ dynamical
symmetry model we will restrict ourselves
to seniority $v=0$ (and consequently reduced isospin $t=0$),
though states with broken nucleon pairs can be easily incorporated
in the formalism.
We begin by specifying the Cartan generators of $so(5)$:
\begin{eqnarray*}
h_{1,i}&=&
\frac{1}{2}(p^\dag_i p_i+p^\dag_{\bar\imath}p_{\bar\imath})-
\frac{1}{2}(n^\dag_i n_i+n^\dag_{\bar\imath}n_{\bar\imath}),\\
h_{2,i}&=&
\frac{1}{2}\sum_{\rho=p,n}(\rho^\dag_i\rho_i+
\rho^\dag_{\bar\imath}\rho_{\bar\imath})-1,
\end{eqnarray*}
where $\rho$ labels protons ($p$) or neutrons ($n$), $i$ is the
$i$-th copy of $so(5)$ and $\bar\imath$ is the time reversal of the
state $i$. We identify $h_1$ with the third component
of the isospin and $h_2$ with the nucleon number.

The positive-root vectors are the $T=1$ creation operators
$\{b^\dag_{-1,i}\equiv n_i^\dag n_{\bar\imath}^\dag,
b^\dag_{0,i}\equiv
(n_i^\dag p_{\bar\imath}^\dag+p_i^\dag n_{\bar\imath}^\dag)/\sqrt{2},
b^\dag_{+1,i}\equiv p_i^\dag p_{\bar\imath}^\dag\}$
plus the isospin-raising operator
$T_{+,i}\equiv(p_i^\dag n_i+p_{\bar\imath}^\dag n_{\bar\imath})/\sqrt{2}$.
The simple-root vectors are $\{b^\dag_{-1,i},T_{+,i}\}$
and $b^\dag_{-1,i}$ is the singular-root vector~\cite{Ush94}.
These, together with the conjugate operators, close the $so(5)$ algebra.
The isospin subalgebra $su_T(2)\subset so(5)$ is generated by
the $so(5)$ elements that are not related to $b^\dag_{-1,i}$.
That is, $su_T(2)$ is generated by $T_{+,i}$,
$T_{-,i}=(T_{+,i})^\dag$,
and $[T_{+,i},T_{-,i}]=T_{0,i}=h_{1,i}$.

Since $so(5)$ is a rank-two algebra, there are two types of spectral
parameters: $e_{1,\alpha}$ and $e_{2,\beta}$ which shall be
denoted as $w_\alpha$ and $e_\beta$. The upper bounds $M1$ and
$M2$ for the indices $\alpha$ and $\beta$, respectively, are
related to the isospin $T$ and the total number of pairs $N$ via
the expressions $M1=N-T$ and $M2=N$. The scalar products of the
simple roots are $\pi2\cdot\pi2=2$, $\pi1\cdot\pi1=1$, and
$\pi2\cdot\pi1=-1$. In the spherical shell model protons and
neutrons occupy single-particle states with quantum numbers
$(j,m)$. The index $i$ then corresponds to $jm$ and ${\bar\imath}$ to
$j\bar m$. Alternatively, due to the rotational
symmetry, the angular momentum $j$ can be used as a label instead
of $i$ but then the corresponding degeneracy $\Omega_j=(2j+1)/2$
should be taken into account. Finally, the weights $\Lambda_i$ are
the same for all $i$ and are those of the fundamental
representation of $so(5)$, that is, $\Lambda\cdot\pi1=0$ and
$\Lambda\cdot\pi2=1$. Inserting these in Eq.~(\ref{gReq}),
together with the choice $\rho1=0$, $\rho2=-1/g$, and
$z_i=2\varepsilon_i$, we obtain the generalized Richardson
equations for the $T=1$ pn-pairing:
\begin{eqnarray}
\frac{1}{g}&=&
\sum_{j=1}\frac{\Omega_j}{2\varepsilon_j-e_\alpha}+
\sum_{\beta(\neq\alpha)}^N\frac{2}{e_\alpha-e_\beta}+
\sum_{\gamma=1}^{N-T}\frac{1}{w_\gamma-e_\alpha},
\nonumber\\
0&=& \sum_{\alpha=1}^N\frac{1}{e_\alpha-w_\gamma}+
\sum_{\delta(\neq\gamma)}^{N-T}\frac{1}{w_\gamma-w_\delta}.
\label{Links eqs.}
\end{eqnarray}
The particular case of $\Omega_j=1$ was derived
by Links {\it et al.}~\cite{Lin02} using the algebraic Bethe ansatz.
Each solution of the equations~(\ref{Links eqs.})
gives an eigenstate of the pn-pairing Hamiltonian
\begin{equation}
H_{\rm pn}=\sum_{jm}2\varepsilon_j(h_{2,jm}+1)-
g\sum_{\mu,jm}b^\dag_{\mu,jm}b_{\mu,jm}, \label{HO5}
\end{equation}
with eigenvalues $E=\sum_{\alpha=1}^N e_\alpha$.
The spectral parameters $e$ are interpreted
as pair energies as in the case of $su(2)$ pairing.
However, due to the larger rank of the $so(5)$, a new
set of spectral parameters $w_\delta$ appears in the
equations~(\ref{Links eqs.}). These new parameters $w$ are associated with
the $su(2)_T$ isospin subalgebra. For each possible isospin $T$
there are $N-T$ new $w$-parameters. The meaning of this new set of
parameters $w$, which do not appear in the expression for the eigenvalues,
becomes evident
when analyzing the eigenstates of~(\ref{HO5}). The Bethe
{\it ansatz} for the $so(5)$ eigenstates of the RG model is a
factorized product wave function. It consists, for
$N<\sum_j\Omega_j$, of a neutron-pair product state related to
the lowest-weight state of $su_T(2)$ and a isospin-raising
product operator state that guaranties the required total isospin
of the ansatz:
\begin{eqnarray*}
\left| w,e;\varepsilon \right\rangle &=&
\prod_{\gamma=1}^{N-T}T_+(w_\gamma)
\prod_{\alpha=1}^Nb_{-1}^+(e_\alpha)
\left|{\rm o}\right\rangle,
\end{eqnarray*}
where the spectral dependence of the generators is
given by the expressions
$$
T_+(w_\gamma)=\sum_i\frac{T_{+,i}}{2\varepsilon_i-w_\gamma},
\quad
b_\mu^\pm(e_\alpha)=\sum_i\frac{b^\pm_{\mu,i}}{2\varepsilon_i-e_\alpha}
$$
Note that application of $T_+$ on $b_{-1}^\dag$ results in $b_0^\dag$,
and application of $T_+$ on $b_0^\dag$ results in $b_{+1}^\dag$
which provides us with Ushveridize's
linear combination of powers of $b_\mu^\dag$~\cite{Ush94}.


Further insight into the structure of the equations~(\ref{Links eqs.})
can be gained by using the classical electrostatic analogy~\cite{Elec}.
They can be considered as the
equilibrium condition for a 2D classical electrostatic problem
involving three sets of charged particles: (a) the orbitons with
charges $-\Omega_j/2$ and fixed positions $2\varepsilon_j$;
(b) the pairons with unit positive charges and free positions
$e_\alpha$; and (c) the new $w$ particles with unit positive charges
and free positions $w_\delta$. In addition, there is a uniform
electric field in the vertical direction (the real axes) with
strength $\sim1/g$. Furthermore, the $w$ particles do not interact
with the orbitons and do not feel the electric field, which can bee seen from the second equation in (\ref{Links eqs.}).

As an example, we performed numerical calculations for $^{64}$Ge
(4 protons and 4 neutrons)
using the single-particle energies (in MeV) $\epsilon_{3/2}=0.00$,
$\epsilon_{5/2}=0.77$, $\epsilon_{1/2}=1.11$, and
$\epsilon_{9/2}=3.00$ and two pn-pairing strengths,
$g=0.1$ (weak) and $0.37$ (strong). We first checked,
in this non-trivial case, that the solutions of~(\ref{Links eqs.})
are indeed the exact solution of the Hamiltonian~(\ref{HO5}) by
direct comparison with a standard nuclear shell-model calculation.

\begin{figure}[htbp]
\vspace{-0.3cm} \hspace{-.8cm}
\includegraphics[width=9.3cm]{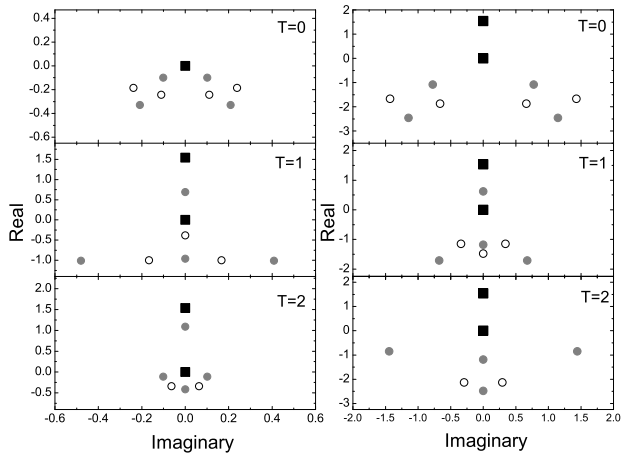}
\caption{Two-dimensional representation of
the $e$ pairon and $w$ particle positions in $^{64}$Ge for the
lowest-energy states with isospin $T=0$, $T=1$, and $T=2$. The left
panel corresponds to weak coupling $g=0.1$ and the right panel to
strong coupling $g=0.37$. The black squares represent the two
lowest orbitons ($p_{3/2}$, $f_{5/2}$) with positions ($2\varepsilon_j$),
the grey circles are the $e$ pairon positions,
and the open circles are $w$ particle positions.}
\label{T2}
\end{figure}

Figure~\ref{T2} shows the solutions for the lowest $T=0$, 1, and 2
states. The $T=0$ solution corresponds to the ground state, while
the $T=1$ and $T=2$ solutions are excited states in $^{64}$Ge. As
in the $su(2)$ pairing case, the different configurations can be
classified in the weak-coupling limit. As can be seen from
Fig.~\ref{T2}, at weak coupling four pairs occupy the $p_{3/2}$
(the four pairons are close to the $p_{3/2}$) in the T=0 state.
This configuration is not allowed for $T=1$ and $T=2$ states due
to the Pauli principle. Correspondingly, three pairons are close
to the lowest $p_{3/2}$ orbiton and the fourth pairon approaches
the orbiton $f_{5/2}$. In all cases the $w$ particles are
intertwined with the pairons to minimize the Coulomb interaction.
The number of $w$ particles ($N-T$), together with the initial
configuration at weak coupling, defines each eigenstate of the
pn-pairing Hamiltonian. As $g$ increases, the free charges expand
under the influence of the electric field and their mutual
interaction. The solutions are subject to numerical instabilities
due to singularities arising when a real pair energy $e$ crosses a
single-particle energy or when real $e$ and $w$ parameters cross.
An example of the first class of crossings can be observed in
Fig.~\ref{T2} for $T=2$ where the pairon associated with the
$f_{5/2}$ orbiton at weak coupling goes down with increasing $g$
and crosses the $p_{3/2}$ orbiton. The $T=1$ case shows an
exchange of positions in the real axis of a pairon and a $w$
particle as an example of the second class of singularities.  The
first class of singularities were already present in the $su(2)$
pairing case and they precluded the practical use of the exact
solution for a long time.  Recently, a new method to overcome this
numerical problem was proposed~\cite{Romb}. We believe that
the same procedure can be used to treat the second class of
singularities as well, making feasible the exact solution of the
$so(5)$ model for very large systems.

In summary, we have presented the generalization of
RG models to arbitrary semi-simple Lie algebras, which include most of the
dynamical-symmetry models of nuclear physics. The generalized
RG models allow the introduction of one-body symmetry breaking
terms like non-degenerate single-particle energies. As an example
of this approach, we gave the exact solution of the $so(5)$
pn-pairing model.
We emphasize
that the exact solution for large systems with $so(5)$ symmetry
could be of great importance in condensed-matter physics
in addressing the phenomenon
of high $T_{\rm c}$-superconductivity~\cite{Z1,S1}.
Finally, the treatment of higher-rank algebras like $sp(6)$ and $so(8)$
opens the possibility of exact nuclear structure calculations
with more realistic quantum integrable models.

This work was supported in part by the Spanish
DGI under grant No.~BFM2003-05316-C02-02 and by the CICYT-IN2P3
cooperation. V.G.G. acknowledges full financial support provided
by a fellowship from the NATO scientific program.

\end{document}